# Sign Language-Based versus Touch-Based Input for Deaf Users with Interactive Personal Assistants in Simulated Kitchen Environments

PAIGE DEVRIES[*]

Gallaudet University, paige.devries@gallaudet.edu

NINA TRAN

Gallaudet University, ninatktran@gmail.com

KEITH DELK

Gallaudet University, keith.delk@gallaudet.edu

MELANIE MIGA

University of Connecticut, melanie.miga@uconn.edu

RICHARD TAULBEE

Gallaudet University, richard.taulbee@gallaudet.edu

PRANAV PIDATHALA

University of Maryland, ppidatha@terpmail.umd.edu

ABRAHAM GLASSER

Gallaudet University, abraham.glasser@gallaudet.edu

RAJA KUSHALNAGAR

Gallaudet University, raja.kushalnagar@gallaudet.edu

CHRISTIAN VOGLER

Gallaudet University, christian.vogler@gallaudet.edu

In this study, we assess the usability of interactive personal assistants (IPAs), such as Amazon Alexa, in a simulated kitchen smart home environment, with deaf and hard of hearing users. Participants engage in activities in a way that causes their hands to get dirty. With these dirty hands, they are tasked with two different input methods for IPAs: American Sign Language (ASL) in a Wizard-of-Oz design, and smart home apps with a touchscreen. Usability ratings show that participants significantly preferred ASL over touch-based apps with dirty hands, although not to a larger extent than in comparable previous work with clean hands. Participants also expressed significant enthusiasm for ASL-based IPA interaction in Netpromoter scores and in questions about their overall preferences. Preliminary observations further suggest that having dirty hands may affect the way people sign, which may pose challenges for building IPAs that natively support sign language input.

---

[*] Nina Tran and Paige DeVries contributed equally to this paper.

## 1 INTRODUCTION

Sign languages exist worldwide and make up an important part of communication for those who are deaf, hard of hearing, or have hearing loss [40]. In the US, around 500,000 people report using American Sign Language (ASL) as their primary mode of communication [30]. Interactive Personal Assistants (IPAs) currently offer voice, app and touch-based interaction options [2][3][26]. Notably, they do not support sign language interaction, which reduces their accessibility and usability for many deaf and hard of hearing users. Although many sign language users are bilingual [19], they are frequently limited in their ability to make themselves understood to an automatic speech recognition system (ASR) due to their non-standard speech patterns [13][24]. Thus, there is a continued need to investigate ways to integrate sign language into IPAs, despite numerous technical challenges [9]. This pairs well with world-wide community-sourced efforts, such as the ASL Citizen Project [1], to build datasets to further the capabilities of sign language recognition technologies to enhance the user experiences of IPAs.

Prior work has investigated how ASL could feature in IPA technology through Wizard-of-Oz experiments [14][15][16][38][33]. This paper is a sequel to recently published work that compared the usability and user preferences in ASL, smart home apps, and Tap-to-Alexa in a Wizard-of-Oz style experiment set in a living room, with a particular focus on how ASL vocabulary and grammar are expressed in such interactions [37]. That work found that usability measures were not significantly different across these three input methods, even though users expressed a slight preference for ASL. Those results came as a surprise – we had expected to see significant usability advantages for ASL. In this study, we follow up with the question if having dirty hands would change the usability picture in favor of ASL over touchscreen-based options. To this end, we explore IPAs in a simulated kitchen environment, which constitutes an important and popular use case [22]. In contrast to earlier work, a kitchen setting necessitates participants getting their hands dirty.

Having dirty hands may affect the way users produce sign language, which a future recognition system built into IPAs would need to consider, so as to be usable in kitchen settings. Additionally, dirty hands may impact interacting with an Alexa device or other IPA through text or touch-based input methods, and potentially favor ASL-based interaction methods over alternatives more than previous work has found. To our knowledge, this is the first study that directly investigates user behavior and preferences with dirty hands while interacting with IPAs via ASL and touch.

The experimental design in this paper is driven by the following two research questions:

- **RQ1:** What are deaf and hard of hearing users' preferences for interacting with IPAs in settings that require them to get their hands dirty?
- **RQ2:** In what ways do dirty hands affect sign language production when interacting with IPAs?

This paper provides some answers to RQ1. Although these answers may seem obvious on the face of it, prior experiments have shown that our hypotheses on ASL usability vs touch-based usability with IPAs have not been confirmed [37], and it was necessary to test this question explicitly in an experiment.

We do not answer RQ2 directly in this paper but have collected the necessary video recordings in the underlying experiment toward answering it. A full answer will require annotating the recordings and analyzing the annotations, which is a time-consuming process. This analysis is planned for future work. In the following we describe related work, the experimental setup and methods, quantitative results on the usability of IPAs with dirty hands and conclude with a discussion of the implications.

## 2 BACKGROUND AND RELATED WORK

IPAs are increasingly mainstream, found in people's pockets, homes, offices, and vehicles. Despite their ubiquity, their accessibility for the deaf and hard of hearing (DHH) community – particularly in the kitchen – remains limited. The kitchen is the second most common space following the living room in a household where IPAs are placed [23][22] and provokes the risk of soiled or dirty hands. The kitchen is also a space of social gathering where deaf people often interact with each other due to the nature of its open access, which facilitates unobstructed visual communication essential for sign language [21]. While the existing literature extensively explores the use of IPAs and advancements in sign language recognition technologies, there remains a notable gap in research addressing deaf people's experiences using IPAs in kitchen settings. Prior work largely focuses on usability and accessibility of IPAs for the deaf community, however not in examining the practical applications and efficacy of interfaces in specific domestic contexts [9][13][38].

Li et. al [25] details how IPAs for DHH users must transcend traditional auditory commands and provide tactile or visual interfaces, such as touchscreen-based devices [27] or gesture recognition systems [36][33]. This adaptation is crucial in the kitchen, where tasks demand hands-on interaction and environmental awareness.

IPAs have become ubiquitous in facilitating daily tasks through voice commands for the larger general population [12][13], yet many DHH users still face significant challenges interacting with these devices in their preferred modalities, such as sign language. A major concern involves the accuracy level of the devices in understanding sign language [16]. Recent research indicates a disparity in the usage of IPAs by DHH individuals compared to the hearing population, primarily due to accessibility issues [15]. Prior work noted that DHH users are interested in utilizing IPAs, but they are hindered by the current limitations of these devices to recognize sign language effectively [33][38][28][14][9]. This reveals a significant gap in the market for IPAs that can cater to sign language users. Future advancements in sign language recognition technologies may one day significantly improve the usability of IPAs for sign language users; in the meantime, it is necessary to study how users would interact with IPAs.

Some DHH users may feel comfortable enough to or prefer using their voice. However, speech variability in ASR systems has also been challenging, due to the high variability in DHH users' – frequently dysarthric – speech [13][24][8][11][18]. To counter this and include people with speech difficulties, there are emerging efforts to optimize ASR technology for non-standard speech [5][17][20], and cross-company consortium work like the Speech Accessibility Project at the Beckman Institute for Advanced Science and Technology.

## 3 RESEARCH METHODS

The study employed a within-subjects repeated measures experimental design with two conditions to compare the usability of ASL with Alexa on an Echo Show device, and iPad-based smart home apps with Alexa ("Apps"). The ASL condition employed a Wizard-of-Oz setup where an interpreter acted behind the scenes via an audio and video link to relay the participants' ASL as voice commands to Alexa. The interactions were set up in the limited domain of a simulated kitchen environment. These are unique in that people in them often have dirty hands, which we simulated through participants creating toy slime with their hands while following instructional "recipe" videos and interacting with Alexa. Note that Echo Show devices are not waterproof, and for this reason we did not test the built-in Tap-to-Alexa input method. The study was conducted in both ASL and English, depending on participant preferences for communication and answering questions. Participant interactions with Alexa were in ASL for the first condition and in English for the second one.

### 3.1 Participant Demographics

We recruited 31 participants (22 female, 8 male, 1 transmasculine) for an in-person study. Participants identified as White (19), Black (5), Asian (2), and other (4). The mean age was 36.5, ranging from 18-64. Most (77.4%) identified as Deaf, 22.6% as hard of hearing, and 6.4% as other. About 87.1% preferred to communicate in ASL only, while the rest used Pidgin Signed English (PSE) and Spoken English. To be included, participants had to self-identify as being fluent in ASL, or report that they use sign language as one of their primary forms of communication. Among the participants, 12 had proficient experience with computers and smart technologies, 11 advanced, 6 some, and 2 expert. Ten owned smart home devices; 4 rarely used them, 2 were frequent users, 2 were occasional, and 1 very frequent. They primarily use these devices for tasks like setting timers, playing music, checking weather, and controlling other smart home devices.

### 3.2 Materials

*3.2.1 Equipment*

Our experimental setup centered around an Amazon Echo Show device, configured to present its responses not only through spoken English but also as captions displayed on the screen in response to user commands. An iPad tablet was used for the apps with Alexa portion of the study, and there were two Philips Hue [31] multicolor lights on the table with the Echo Show (see Figure 1). We attached a camera to the Echo Show device, connecting it to a laptop on our local network for real-time video streaming to support our Wizard-of-Oz approach. Additionally, a second camera on a tripod was set up behind the participants to capture both their signing (from their Echo Show-attached camera) and their inputs along with device responses (from the tripod camera). An additional researcher station was set up in a separate room, but on the same local network, with a laptop connected to a high-quality microphone to relay interpreted ASL commands in voice to Alexa. Further details can be found in Section 3.3.2 and the appendix.

*3.2.2 Slime for Simulating Dirty Hands*

To simulate dirty hands, a major component of the participants' tasks was to create toy slime in a mixing bowl, which is a viscous liquid that sticks to the hands and impedes interaction with touch screens. The recipe for this experiment consisted of measuring out and mixing clear glue with water and liquid starch and adding natural food coloring as desired. In a follow-up task, rock salt was added to the finished slime. Participants were given the option of using nitrile gloves to protect their hands and informed about the full list of ingredients and possible allergens. They also had access to a sink with running water, soap, and towels nearby the Echo Show station, which they could use on an as-needed basis. The IPA equipment and the slime bowl and ingredients were all situated on the same station, so that participants could watch and interact with the IPA while mixing the slime recipes. The sink, soap and towels were located at a nearby station approximately eight feet away from the IPA station.

*3.2.3 In-Experiment Instruments and Post-Experiment Survey*

We utilized several different systems of measurement to determine what the user preferences are between the ASL and Apps input methods: Adjective scale, Netpromoter Score (NPS), and the System Usability Scale (SUS). SUS is applied to assess the usability of a system [10][34] using a 5-point scale ranging from strongly agree to strongly disagree with ten questions. The adjective scale is used to get a high-level assessment of the usability of a system on a 7-point Likert scale from Worst Imaginable to Best Possible. The NPS is a common tool for assessing how likely it is that someone will recommend a system to someone else, where choices along an 11-point scale are classified into promoters, passives, and detractors. The original versions of SUS, adjective scale and NPS are administered in written English form, but in our study, we gave participants the choice between English or ASL versions [7] which provides a psychometrically validated, equivalent, ASL interpretation of the written English version.

All three usability measures were completed twice by each participant, upon completion of each respective condition (ASL and Apps). A post-experiment survey asked participants their feelings on the different input methods, what they preferred, what they might use an IPA for rated in terms of importance, and if they could see themselves using IPAs in a smart home environment.

*3.2.4 Task Lists*

During the study, participants were tasked with following two instructional videos embedded in the Alexa system, in order to simulate watching and following along a recipe video in the kitchen (see Figure 2). They were given two task lists to follow. The first provided instructions for making slime as described in Section 3.2.2, and the second provided instructions for mixing in rock salt after completing the first list. Both task lists required interacting with Alexa to retrieve the instructional videos, control lights and set timers. The order of conditions was counterbalanced, with half of the participants using ASL for the first task and then continuing with Apps for the second, and the other half starting with Apps then continuing with ASL. The appendix contains further details.

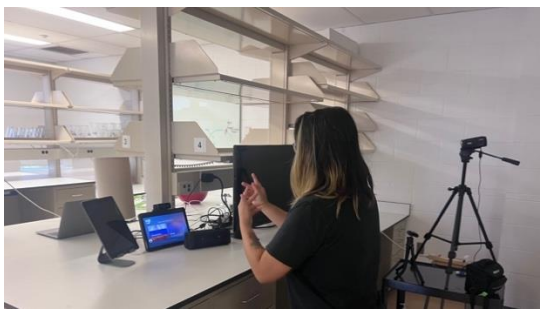 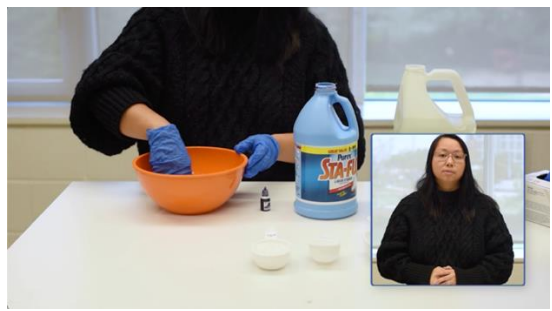

Figure 1: A participant signs to the Echo Show device, with the iPad nearby and the webcam on top feeding video to the ASL interpreter behind the scenes (the "Wizard").

Figure 2: Instructional video of how to create slime sticking to hands. Participants had to copy the steps.

### 3.3 Methods

*3.3.1 Participant Procedures*

Participant sessions were up to one hour in length, with an average duration of 35 minutes. Participants were compensated for their time in the study, which received ethics approval. A deaf researcher fluent in both ASL and English guided each participant through the informed consent and video release procedures in their preferred language. They then completed an intake survey asking their demographic information and experience with IPAs and other smart home technologies. Following the intake survey, the participants positioned themself in front of the Echo Show and began following their first task list. Figure 1 shows a participant signing to Alexa during the ASL condition, and Figure 2 illustrates one of the two instructional videos we created for participants to follow on making slime. Note that the instructional videos had versions with and without captions, and participants were able to select their preferred version via an Alexa command.

The Apps with Alexa condition instructed participants to use an iPad to communicate with Alexa, within several different apps that each related to specific commands, the Philips Hue app for the lights, the YouTube app for the instructional video, and the Alexa app for the timer. After completing each, the participant completed the SUS, NetPromoter and Adjective Scale. After completing all conditions and tasks, the participant filled out a post-experiment survey with overall impressions.

*3.3.2 Wizard Procedures*

The experimental design requires that the participant remains unaware of the ASL interpreter (the "Wizard") behind the scenes. This depends on both the visibility of participants on the Echo Show camera to the ASL interpreter in a separate room unbeknownst to the participant and a clear audio connection between the Echo Show and Wizard stations, so that the Echo Show could receive commands from the Wizard. The participants were not aware that the Echo Show-mounted camera was connected to a laptop rather than to the Alexa device itself, which is the assumption they likely came to during the experiment. The intention behind this was to simulate machine understanding of ASL, and if the participants had been aware of this being a simulation, their reactions may have changed, and biases attached to the input methods might have been introduced. For coordination, unbeknownst to the participants, the research team communicated through the Discord app on their phones as needed.

*3.3.3 Data Analysis*

We performed both descriptive and inferential statistical analyses on the SUS and adjective scale scores via paired t-tests for the two conditions. We further analyzed the NPS via the nonparametric Wilcoxon signed-rank test [39]. Additionally, we calculated descriptive statistics for the post-experiment survey questions.

## 4 RESULTS

In the following, we provide the usability results, participant preferences from the post-experiment survey, and findings regarding ASL usage with IPAs in simulated kitchen scenarios.

### 4.1 Usability Results

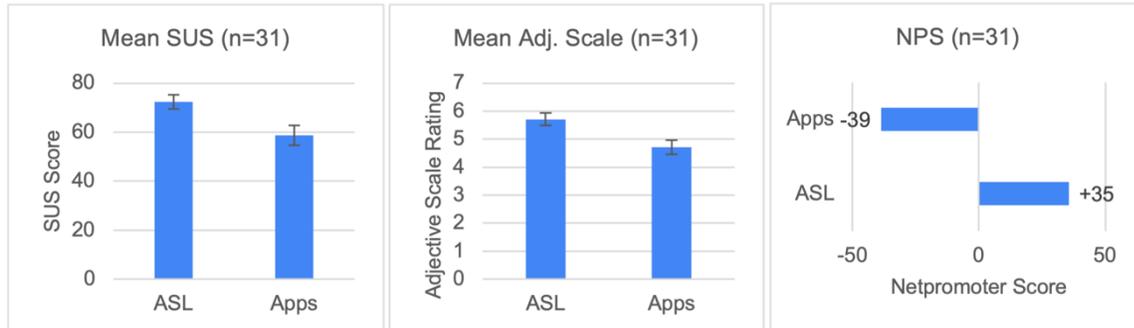

Figure 3: Mean SUS for ASL vs apps with dirty hands. The pairwise difference was statistically significant.

Figure 4: Mean Adjective Scale Rating for ASL vs apps with dirty hands. The pairwise difference was statistically significant.

Figure 5: NPS for apps vs ASL with dirty hands. Apps were rated unfavorably, and ASL was rated favorably. The difference was statistically significant.

Figure 3 shows the mean SUS for the ASL and App conditions. ASL was preferred, with a mean SUS of 72.4 (SD 16.2, SE 2.906), versus Apps with a mean SUS of 58.8 (SD 22.5, SE 4.034). Similarly, Figure 4 shows the mean adjective scale ratings, with 1 equaling the worst rating and 7 equaling the best rating. ASL was preferred there, too, with a mean rating of 5.7 (SD 1.2, SE 0.218) vs Apps with a mean rating of 4.7 (SD 1.4, SE 0.255). Post-hoc paired t-testing was statistically significant for both SUS (df=30, $p<0.005$) and the adjective rating scale (df=30, $p<0.001$).

Figure 5 shows the Netpromoter scores for the two conditions. Apps rated unfavorably with a score of -39 (18 detractors, 6 promoters), while ASL rated favorably, with a nearly opposite score of +35 (5 detractors, 16 promoters). The Wilcoxon-signed rank test showed the difference to be statistically significant ($p<0.001$).

The ASL SUS of 72.4 is considered an "OK" level of usability [6][10][34], while the Apps score of 58.8 falls into the low-to-medium end of marginal usability, below the threshold of what is considered acceptable for a system. These scores are very similar for those from the paper without dirty hands, which this experiment builds on (ASL: 71.6, Apps 56.3) [37]. On the adjective scale, the results are considered "good" for ASL and "OK" for Apps [6]. The Netpromoter score of -39 for Apps is what we would expect with the corresponding SUS for Apps; however, the +35 score for ASL is much higher than the score of -1 that the average SUS in Figure 3 would predict [10][35]. In other words, participants expressed considerable enthusiasm for the ASL to Alexa option, which is not reflected in the SUS.

### 4.2 Post-Experiment Survey Results

When faced with dirty hands, 84% of participants expressed that they favored ASL, while 9.6% favored Apps and 6.4% favored neither. When asked how difficult it was to use Alexa, 64.5% of participants perceived Apps as challenging, while only 16.1% found ASL difficult. A further 19.3% indicated that neither was challenging. Participants were asked to rate their level of interest in using IPAs if they could recognize ASL, with over 85% expressing high interest, and the rest expressing some interest.

Some participants mentioned challenges with Alexa and an iPad, stating, *"Typing with Alexa was cumbersome and made the iPad dirty. On the other hand, signing was a little difficult as the slime stuck to my hands"* (P39). Additionally, some participants mentioned technical difficulties, stating, *"The system did not recognize my signs and it stopped playing the videos several times and even though I signed correctly"* (P29). Other participants were concerned about how misunderstandings can impact their homes when signing something that could be interpreted as a command by an IPA. Preliminary observations from the video recordings showed that participants had to be careful in their signing with dirty hands; for example, one participant splashed themselves while signing. While a full analysis will need to be conducted in future work to answer RQ2, P39's comment and these observations indicate that dirty hands do influence ASL production.

## 5 DISCUSSION

The results of this experiment parallel those from the experiment that this paper builds on, with very similar SUS for ASL and Apps, although in this experiment the difference was statistically significant, demonstrating a preference for the ASL condition. This is despite the design change of having dirty hands. We had expected, with the addition of dirty hands, that scores would be lower across the board, with a larger gap between ASL and Apps than before. There was also a discrepancy between the SUS and NPS results – particularly, the usability of ASL from SUS would have predicted a neutral response in the NPS. Instead, the NPS reflected considerable enthusiasm for the ASL input method. We see three possible explanations for these skewed results:

- The novelty effect of ASL in IPAs and participants being enthusiastic about this technology becoming a reality, despite usability flaws.
- The contrast in usability between Apps with Alexa and ASL to Alexa, with ASL looking comparatively much better to participants – however, the adjective scale results do not show such a thing with a one-point difference.
- Participants might have been unsure about how to answer some of the ten questions in SUS in the context of the simulated kitchen setting.

We consider a combination of the first and third reason to be the most likely explanation for the observed outcome. This is because most participants had never used an IPA, as it was their first time using one in the study. Few participants faced challenges in understanding the meaning of the SUS questions, with the ASL versions and the researchers present to provide clarification.

An interesting observation is that some participants opted to use nitrile gloves, yet they would remove specifically when signing a command to Alexa. This behavior is likely attributed to the hindrance caused by gloves, particularly when trying to fingerspell the wake word "Alexa" or sign a name sign for Alexa. We also tentatively observed that some participants would alter their signing style due to dirty hands, or get annoyed when they splashed slime over themselves while signing. Overall, the results in this paper suggest that contrary to our expectations, the presence of dirty hands did not materially alter the usability fundamentals of ASL relative to Apps, even though it has now been confirmed that ASL has better usability than Apps, in contrast to earlier work. Furthermore, there remains a considerable need – and some enthusiasm – for options to sign to IPAs. Finally, the exact way ASL is expressed with dirty hands will need further analysis from the recorded data set. Name signs and fingerspelling are harder to execute with dirty hands, which raises the possibility of the letters/signs being formed differently in a way that could be challenging for IPAs.

### 5.1 Limitations

The Wizard-of-Oz setup created several constraints in the user experience that likely would be absent in an IPA that natively supports ASL. For instance, several participants expressed a desire to have the ability to view themselves on screen in the same way they appear to the Alexa device. The absence of this feature leaves them uncertain about whether Alexa can see and understand what they sign, forcing them to rely on trial-and-error testing. This issue was particularly evident with taller participants not within the camera's field of view. Another notable limitation was Alexa misinterpreting the term "rock salt," as voiced by the wizard, triggering responses related to "rock songs" or similar-sounding phrases. The interpretation step introduces considerable chances for error. Finally, Alexa's handling of captioned versus uncaptioned videos, as per user commands, was implemented via custom skills. This posed significant challenges that made it difficult to pause a video in the middle of something to issue another command to Alexa and then resume where the user left off. Timers worked, but the participants had to refrain from several other types of commands while watching.

### 5.2 Future Work

The immediate future avenue for research is to study in detail how dirty hands affect signing style and fingerspelling. To the best of our knowledge, this topic has not yet been mentioned in papers related to automatic sign language recognition. Additionally, future work should build off this work by examining the user preferences deaf individuals report when engaging with other popular IPAs, such as Apple's Siri and Google's Home and Nest products, paying special attention to how smoothly (or not) requesting a specific video to play proves to be. Other IPA-supported environments also need to be explored, such as smart cars. Finally, it is important to also consider options for users who do not know sign language and have dysarthric speech that prevents them from using IPA's built-in ASR.

### ACKNOWLEDGMENTS

The contents of this paper were developed under a grant from the National Institute on Disability, Independent Living, and Rehabilitation Research (NIDILRR grant number 90REGE0013). NIDILRR is a Center within the Administration for Community Living (ACL), Department of Health and Human Services (HHS). The contents of this paper do not necessarily represent the policy of NIDILRR, ACL, HHS, and you should not assume endorsement by the Federal Government. Additional funding was provided by a National Science Foundation REU Site Grant (#2150429). Norman Williams supported the technical setup of the experiment. Matthew Seita and James Waller helped with the statistical analysis.

### REFERENCES

[1] Aashaka Desai, Lauren Berger, Fyodor O. Minakov, Vanessa Milan, Chinmay Singh, Kriston Pumphrey, Richard E. Ladner, Hal Daumé, Alex X. Lu, Naomi Caselli and Danielle Bragg. 2023. ASL Citizen: A Community-Sourced Dataset for Advancing Isolated Sign Language Recognition. arXiv:2304.05934v2 [cs.CV]. https://doi.org/10.48550/arXiv.2304.05934

[2] Amazon. [n.d.]. What is Tap to Alexa? https://www.amazon.com/b?ie=UTF8&node=21213735011

[3] Amazon. [n.d.]. Alexa App. https://www.amazon.com/Alexa-App/b?ie=UTF8&node=18354642011

[4] Apple. [n.d.]. Photo Booth. https://apps.apple.com/us/app/photo-booth/id1208226939

[5] Apple. [n.d.] Improved Speech Recognition for People Who Stutter. Retrieved from https://machinelearning.apple.com/research/speech-recognition

[6] Aaron Bangor, Philip Kortum and James Miller. 2009. Determining what Individual SUS Scores Mean: Adding an Adjective Rating Scale. Journal of Usability Studies 4(3), 114–123.

[7] Larwan Berke, Matt Huenerfauth and Kasmira Patel. 2019. Design and Psychometric Evaluation of American Sign Language Translations of Usability Questionnaires. ACM Transactions on Accessible Computing (TACCESS), 12(2), 1-43.

[8] Jeffrey P. Bigham, Raja Kushalnagar, Ting-Hao Kenneth Huang, Juan Pablo Flores, and Saiph Savage. 2017. On how deaf people might use speech to control devices." In Proceedings of the 19th International ACM SIGACCESS Conference on Computers and Accessibility, pp. 383-384.

[9] Danielle Bragg, Oscar Koller, Mary Bellard, Larwan Berke, Patrick Boudreault, Annelies Braffort, Naomi Caselli, Matt Huenerfauth, Hernisa Kacorri, Tessa Verhoef, Christian Vogler, and Meredith Ringel Morris. 2019. Sign Language Recognition, Generation, and Translation: An Interdisciplinary Perspective. In Proceedings of the 21st International ACM SIGACCESS Conference on Computers and Accessibility (ASSETS '19). Association for Computing Machinery, New York, NY, USA, 16–31. https://doi.org/10.1145/3308561.3353774

[10] John Brooke. 2013. SUS: A Retrospective. Journal of Usability Studies 8(2), 29–40.

[11] Raymond Fok, Harmanpreet Kaur, Skanda Palani, Martez E. Mott, and Walter S. Lasecki. 2018. Towards More Robust Speech Interactions for Deaf and Hard of Hearing Users. In Proceedings of the 20th International ACM SIGACCESS Conference on Computers and Accessibility (ASSETS '18). Association for Computing Machinery, New York, NY, USA, 57–67. https://doi.org/10.1145/3234695.3236343

[12] Abraham Glasser, Kesavan Kushalnagar, and Raja Kushalnagar. 2017. Deaf, hard of hearing, and hearing perspectives on using automatic speech recognition in conversation. Proceedings of the 19th International ACM SIGACCESS Conference on Computers and Accessibility.

[13] Abraham Glasser. 2019. Automatic Speech Recognition Services: Deaf and Hard-of-Hearing Usability. In Extended Abstracts of the 2019 CHI Conference on Human Factors in Computing Systems (CHI EA '19). Association for Computing Machinery, New York, NY, USA, Paper SRC06, 1–6. https://doi.org/10.1145/3290607.3308461

[14] Abraham Glasser, Vaishnavi Mande, and Matt Huenerfauth. 2021. Understanding Deaf and Hard-of-Hearing Users' Interest in Sign-Language Interaction with Personal-Assistant Devices. In Proceedings of the 18th International Web for All Conference (W4A '21), April 19-20, 2021, Ljubljana, Slovenia. ACM, New York, NY, USA, 11 pages. https://doi.org/10.1145/3430263.3452428

[15] Abraham Glasser, Matthew Watkins, Kira Hart, Sooyeon Lee, and Matt Huenerfauth. 2022. Analyzing Deaf and Hard-of-Hearing Users' Behavior, Usage, and Interaction with a Personal Assistant Device that Understands Sign-Language Input. In Proceedings of the 2022 CHI Conference on Human Factors in Computing Systems (CHI '22). Association for Computing Machinery, New York, NY, USA, Article 306, 1–12. https://doi.org/10.1145/3491102.3501987

[16] Abraham Glasser. 2023. Empirical Investigations and Dataset Collection for American Sign Language-Aware Personal Assistants. Available from ProQuest Dissertations & Theses Global. (2796329430).

[17] Google. 2023. Project Euphonia. Retrieved from https://sites.research.google/euphonia/about/

[18] Linda Gottermeier, and Raja Kushalnagar. 2016. User evaluation of automatic speech recognition systems for deaf-hearing interactions at school and work. Audiology Today 28.2 (2016): 20-34.

[19] Francois Grosjean. 2010. Bilingualism, biculturalism, and deafness. International Journal of Bilingual Education and Bilingualism, 13(2), pp.133-145.

[20] Larry Hardesty. 2021. Voiceitt Extends The Voice Revolution To People With Nonstandard Speech. (June 2021). Retrieved January 4, 2024 from http://ccrma.stanford.edu/~jos/bayes/bayes.html    https://www.amazon.science/latest-news/voiceitt-extends-the-voice-revolution-to-people-with-nonstandard-speech

[21] Charlene A. Johnson. 2010. Articulation of Deaf and Hearing Spaces Using Deaf Space Design Guidelines: A Community Based Participatory Research with the Albuquerque Sign Language Academy. https://digitalrepository.unm.edu/arch_etds/18/

[22] Valerie K. Jones. 2022. Why people use virtual assistants: Understanding engagement with Alexa. Journal of Brand Strategy, 11(1), pp.80-101.

[23] Bret Kinsella and Ava Mutchler. 2018. Smart Speaker Consumer Adoption Report. Voicebot.ai. https://voicebot.ai/wp-content/uploads/2018/10/voicebot-smart-speaker-consumer-adoption-report.pdf

[24] Kashyap Kompella. 2023. Recognizing Atypical Speech Is ASR's Achilles' Heel. Speech Technology, vol. 28, no. 3, p. 8.

[25] Franklin Mingzhe Li, Jamie Dorst, Peter Cederberg, and Patrick Carrington. 2021. Non-Visual Cooking: Exploring Practices and Challenges of Meal Preparation by People with Visual Impairments. In Proceedings of the 23rd International ACM SIGACCESS Conference on Computers and Accessibility (ASSETS '21). Association for Computing Machinery, New York, NY, USA, Article 30, 1–11. https://doi.org/10.1145/3441852.3471215

[26] Irene Lopatovska, Katrina Rink, Ian Knight, Kieran Raines, Kevin Cosenza, Harriet Williams, Perachya Sorsche, David Hirsch, Qi Li, and Adriana Martinez. 2018. Talk to me: Exploring user interactions with the Amazon Alexa. Journal of Librarianship and Information Science, 51(4), 984–997. https://doi.org/10.1177/0961000618759414

[27] Michal Luria, Guy Hoffman, and Oren Zuckerman. 2017. Comparing Social Robot, Screen and Voice Interfaces for Smart-Home Control. In Proceedings of the 2017 CHI Conference on Human Factors in Computing Systems (CHI '17), 580--592. https://dl.acm.org/doi/10.1145/3025453.3025786

[28] Vaishnavi Mande, Abraham Glasser, Becca Dingman, and Matt Huenerfauth. 2021. Deaf Users' Preferences Among Wake-Up Approaches during Sign-Language Interaction with Personal Assistant Devices. In Extended Abstracts of the 2021 CHI Conference on Human Factors in Computing Systems (CHI EA '21). Association for Computing Machinery, New York, NY, USA, Article 370, 1–6. https://doi.org/10.1145/3411763.3451592

[29] Campbell McDermid. 2018. Learning to Interpret: Working from English into American Sign Language. RIT Press.

[30] Ross E. Mitchell, Travas A. Young, Bellamie Bachleda, and Michael A. Karchmer. 2005. How Many People Use ASL in the United States? Why Estimates Need Updating. In Sign language Studies, Volume 6, Number 3, 2006. 37 pages.

[31] Philips. [n.d.]. The Philips Hue app. https://www.philips-hue.com/en-us/explore-hue/apps/bridge

[32] RealVNC.. [n.d.] VNC Viewer. https://www.realvnc.com/en/connect/download/viewer/

[33] Jason Rodolitz, Evan Gambill, Brittany Willis, Christian Vogler, and Raja Kushalnagar. 2019. Accessibility of Voice-Activated Agents for People who are Deaf or Hard of Hearing. Journal on Technology and Persons with Disabilities 7 (2019), 144–156. http://hdl.handle.net/10211.3/210397

[34] Jeff Sauro and James R. Lewis. 2011. When designing usability questionnaires, does it hurt to be positive? In Proceedings of the SIGCHI Conference

on Human Factors in Computing Systems (CHI '11). Association for Computing Machinery, New York, NY, USA, 2215–2224. https://doi.org/10.1145/1978942.1979266

[35] Jeff Sauro. 2012. Predicting Net Promoter Scores from System Usability Scale Scores. MeasuringU. https://measuringu.com/nps-sus/ (last accessed: January 25, 2024)

[36] Julia Schwarz, Charles Claudius Marais, Tommer Leyvand, Scott E. Hudson, and Jennifer Mankoff. 2014. Combining body pose, gaze, and gesture to determine intention to interact in vision-based interfaces. In CHI '14: Proceedings of the SIGCHI Conference on Human Factors in Computing Systems, April 26-May 1, 2014, Toronto, Canada, 9 pages. https://doi.org/10.1145/2556288.2556989

[37] Nina Tran, Paige DeVries, Matthew Seita, Raja Kushalnagar, Abraham Glasser and Christian Vogler. 2024. Assessment of Sign Language-Based versus Touch-Based Input for Deaf Users Interacting with Intelligent Personal Assistants. In Proceedings of the 2024 CHI Conference on Human Factors in Computing Systems (CHI 2024). https://doi.org/10.1145/3613904.3642094

[38] Gabriella Wojtanowski, Colleen Gilmore, Barbra Seravalli, Kristen Fargas, Christian Vogler, and Raja Kushalnagar. 2020. "Alexa, Can You See Me?" Making Individual Personal Assistants for the Home Accessible to Deaf Consumers. The Journal on Technology and Persons with Disabilities (2020), 130.

[39] Robert F Woolson. 2007. Wilcoxon signed-rank test. Wiley encyclopedia of clinical trials, 1-3.

[40] World Federation of the Deaf. [n.d.]. Our work. http://wfdeaf.org/our-work/

## A APPENDICES

Note that the custom skill to play videos is called "Alice" in honor of Alice Cogswell, a key figure in deaf education in the United States.

### A.1 ASL Task List – Slime

Before starting the task list
1. Give the participant a few practice options:
2. Turn on the lights
3. Change the light color
4. Turn off the lights
5. Go home

Begin the task list
1. "Alexa, turn on the lights."
2. "Alexa, ask Alice to play slime recipe"
3. If you want captions, "Alexa, ask Alice to play slime recipe with captions"
4. Watch the video play and follow the directions.
5. If you want to pause the video, "Alexa, pause."
6. If you want to resume the video, "Alexa, resume."
7. If you miss one of the steps, "Alexa, go back." or "Alexa, skip back."
8. If you want to skip one of the steps, "Alexa, go forward." or "Alexa, skip."
9. If the video disappears, repeat step 2.
10. "Alexa, pause" after food coloring is added.
11. "Alexa, set a timer for 30 seconds."
12. Mix in food coloring until combined.
13. When the timer runs out, "Alexa, stop"
14. After the timer disappears, wait about 10 seconds.
15. "Alexa, resume."
16. Continue watching the video and following the instructions until it ends.

17. When the video ends, "Alexa, go home."
18. "Alexa, change the light color to [match the color of your slime]."
19. "Alexa, dim the lights 50%."
20. "Alexa, turn off the lights."
21. Proceed with the rock salt task (Apps).

### A.2   ASL Task List – Rock Salt

Before starting the task list
1. Give the participant a few practice options:
2. Turn on the lights
3. Change the light color
4. Turn off the lights
5. Go home

Begin the task list
1. "Alexa, ask alice to play rock salt"
2. If you want captions, "Alexa, ask Alice to play rock salt with captions"
3. Watch the video play and follow the directions.
4. If you want to pause the video, "Alexa, pause."
5. If you want to resume the video, "Alexa, resume."
6. If you miss one of the steps, "Alexa, go back." or "Alexa, skip back."
7. If you want to skip one of the steps, "Alexa, go forward."  or "Alexa, skip."
8. If the video disappears, repeat step 1.
9. "Alexa, pause the video" when you need to knead.
10. "Alexa, set a timer for 2 minutes."
11. Knead until it comes together.
12. "Alexa, stop the timer."
13. After the timer disappears, wait about 10 seconds.
14. "Alexa, resume the video."
15. Continue watching the video until it ends.
16. If you miss one of the steps, "Alexa, go back."
17. When the video ends, "Alexa, go home."
18. Start sculpting into whatever shape you want.
19. "Alexa, what's the weather like?"
20. "Alexa, turn off the lights."
21. "Alexa, can I dispose of slime in the trash?"
22. "Alexa, how long do I wash my hands to clean off slime?"
23. "Alexa, go home."

### A.3 Apps Task List - Slime

Before starting the task, provide training on:
1. Where to find each App
2. How to interact with the Alexa App for timer (type), stopping a timer (type)
3. How to interact with the hue app for lights

Begin the task list
1. Open the YouTube App.
2. Look for the video named "Slime Recipe"
3. Watch the video play and follow the directions.
4. Pause the video after food coloring is added.
5. Open the Alexa App.
6. Set a timer for 30 seconds.
7. Mix in food coloring until combined.
8. Stop the timer.
9. Open the Hue App.
10. Turn on the lights.
11. Change the light color to match the color of your slime.
12. Open the YouTube App.
13. Resume the video.
14. Add ¼ cup of liquid starch to the bowl.
15. Continue watching the video until it ends.
16. Open the Hue App.
17. Dim the lights 50%.
18. Proceed with the rock salt task (ASL).

### A.4 Apps Task List – Rock Salt

Before starting the task, provide training on:
1. Where to find each App
2. How to interact with the Alexa App for timer (type), stopping a timer (type)
3. How to interact with the hue app for lights

Begin the task list
1. Open the YouTube App.
2. Look for the video named "Rock Salt"
3. Watch the video play and follow the directions.
4. Pause the video when you need to knead.
5. Open the Alexa App.
6. Set a timer for 2 minutes or longer.
7. Knead until it comes together.

8. Stop the timer.
9. Open the YouTube App.
10. Resume the video.
11. Continue watching the video until it ends.
12. When done, start sculpting into whatever shape you want.
13. Open the Alexa App.
14. Ask Alexa to show you the weather.
15. Open the Hue App.
16. Turn off the lights.
17. Open the Alexa App.
18. Ask Alexa if you can dispose of slime in the trash.
19. Ask Alexa how long to wash your hands to clean off slime.

## A.5 Extended Experimental Procedures

**Equipment**

Nearby the station with the Echo Show (but with the screen angled to be out of sight of the participant) was one of two MacBook Air laptops connected to each other, the Echo Show-mounted webcam, and an EarFun UBOOM 28W speaker that we used in replacement of the laptop's built-in speakers for clearer output. The researcher in the room with the participant did not interact with the laptop to avoid risking the participant becoming aware of the additional connected device.

**Wizard Procedures**

The "Wizard" station consisted of the second MacBook Air laptop connected to the Echo Show-attached camera. Both laptops were also connected to each other through hardwired Ethernet cable for reliable connectivity. The app we used to connect the two computers was FaceTime, connected with the Echo Show-attached webcam. The station also had a Blue Yeti microphone for audio input to the speaker in the Echo Show station. Through VNC viewer [32] and an additional Mac desktop computer used as a second monitor, the Wizard controlled both MacBook Laptops, established and maintained a FaceTime connection during interviews, recorded the participant on the Echo Show-attached webcam through remote controlling of the "Dorothy" laptop and the laptop's Photo Booth app [4]. With all the above-mentioned equipment connected, the Wizard was able to see, record, and interact with the participant through interpretation from ASL to spoken English for machine understanding, as well as hear Alexa's verbal output.

To facilitate communication between deaf participants and Alexa in the ASL to Alexa condition in a manner representative of potential future ASL recognition systems, the Wizard was required to use literal interpretation for each session. This approach contrasts with what interpreter training typically practices, as most other scenarios require interpreters to use functionally/dynamically equivalent interpretation. In the latter situation, interpreters voice the concepts that are implied by what they see [29]. This created tension for the Wizard and required them to be continually monitored by the deaf research team to ensure they did not inadvertently step out of their role. To further minimize possible interpretation biases, the Wizard was only informed of the condition order, but not of the specifics of each task list.

The Wizard specifically looked for the wake word. Participants were allowed to either fingerspell Alexa or use a name sign for Alexa (e.g., "FS(ALEXA)" or "NS(AX);"). If participants omitted the wake word, the Wizard did not speak

Alexa's name, but still interpreted the command produced by the participant. Note that some participants encountered difficulty in signing the wake word and recalling it due to their limited exposure to IPA technology.

If Alexa did not receive a command, it was up to the participant to become aware. If there was no response from Alexa, the Wizard did not repeat the command and instead waited for the participant to decide how to proceed with the task. Some participants chose to repeat the task item, others chose to move on to the next one. Note that the Echo Show has a visual indicator showing whether the wake word has been uttered or when a command is being processed in the form of a blue line at the bottom of the display screen, and its presence or absence could provide clues to the participants. If the Wizard failed to understand the participant's signing, they spoke the command "Alexa, write," which forced the response "I'm sorry, I didn't get that" via audio and captions on the Echo Show. This typically prompted the participant to try their command again.